\documentclass[showpacs,amsmath,nofootinbib,amssymb,epsfig]{revtex4}
\usepackage{graphicx} 
\usepackage{dcolumn}
\usepackage{bm}

\usepackage{color}

\usepackage{amssymb}

\begin{document}

\title{The Cosmological Dynamics of Interacting Logarithmic Entropy Corrected Holographic Dark Energy Model}
\author{F. Darabi}\email{f.darabi@azaruniv.ac.ir} \author{F. Felegary}\email{falegari@azaruniv.ac.ir}
\affiliation{Department of Physics, Azarbaijan Shahid Madani University, Tabriz, 53714-161 Iran}
\author{M. R. Setare}\email{rezakord@ipm.ir}
\affiliation{Department of Science, University of Kurdistan, Sanandaj, Iran}

\date{\today}

\begin{abstract}
We investigate the cosmological dynamics of interacting Logarithmic Entropy Corrected Holographic Dark Energy model with Cold Dark Matter. Fixed points
are determined  and their corresponding cosmological models are presented. Moreover, the dynamical properties of these fixed points are derived. 
\end{abstract}
\maketitle

\section{Introduction}
Recent cosmological and astrophysical data from  the observations of type Ia
supernovae, Cosmic Microwave Background radiation (CMB) and Large Scale Structure
(SSL) have proved that the universe experiences an accelerated expansion phase
\cite{riess}. It is commonly believed that the accelerated expansion have been driven by an enigmatic
energy component with negative pressure so called dark energy (DE). The cosmological
constant is the simplest candidate for dark energy. However, this candidate
suffers from two
major problems, namely the fine-tuning and the cosmic coincidence problems \cite{copeland,weinberg}. The nature of dark energy is unknown, thus there are several suggested models for DE, the well-known of them are Thachyon, K-essence, Phantom, Quintom, chaplygin gas, Quintessence and modified gravity \cite{padman,copeland}.

Recently, a new model of DE within the framework
of quantum gravity,  so called Holographic Dark Energy model (HDE), was suggested
\cite{cohen,li}, and its energy density, based on the holographic principle, was introduced as \cite{susskind}
\begin{equation}\label{1}
\rho_{\Lambda}=3c^{2}M_{p}^{2}L^{-2},
\end{equation}
where $c$ is a numerical constant, $M_{p}$ is the reduced Planck mass and $L$
is the cut-off length. 

In the HDE model, the Bekenstein-Hawking entropy  $(S_{BH}=\frac{A}{4G})$
plays an essential role and is satisfied on the horizon \cite{waldd} (where $A\sim L^{2}$ is the area of horizon). Since this model is connected
to the area of entropy,  any correction in the entropy formula will affect the energy
density of HDE model. Corrections may arise because of the quantum field
theory, thermal and quantum fluctuations in Loop Quantum Gravity (LQG) and
string theory \cite{radicella,ashtekar,waldw}.
One correction to the  entropy is the logarithmic  correction \cite{majhi}
\begin{equation}
S_{BH}=\frac{A}{4G}+\tilde{\alpha}\ln(\frac{A}{4G})+\tilde{\beta}.
\end{equation}
Here $\tilde{\alpha}$ and $\tilde{\beta}$ are two dimensionless constants.
The energy density of the Logarithmic Entropy-Corrected Holographic Dark Energy  (LECHDE) can be
obtained as \cite{wei}
\begin{equation}
\rho_{\Lambda}=3c^{2}M_{p}^{2}L^{-2}+\alpha L^{-4} \ln(M_{p}^{2}L^{2})+\beta
L^{-4}, \label{ECHDE}
\end{equation}
where $\alpha$ and $\beta$ are dimensionless constants,\footnote{The second and third
terms in Eq.(\ref{ECHDE}) are comparable to the first term only when $L$
takes a very small value, because the corrections determined by these terms
are important at early universe. When the universe becomes large, corrections
are ignorable and the Logarithmic entropy-corrected holographic dark energy  reduces to the ordinary holographic dark energy. Therefore, estimation of
the parameters  in the present model of early universe is not an easy task
and needs accurate data from early universe which are not so available and
reliable. Of course, for the ordinary holographic dark energy
 model, the confrontation with present observations have been studied in \cite{s1,li2}.} $c$ is a positive constant and the IR cut-off parameter $L$,  selected to be the radius of the event horizon as measured on the sphere of the horizon, is described as
\begin{equation}
L=ar(t),\label{L}
\end{equation}
\begin{equation}
\int_{0}^{r(t)}\frac{dr}{\sqrt{1-kr^2}}=\frac{R_{h}}{a},
\end{equation}
which yields
\begin{equation}
r(t)=\frac{1}{\sqrt{k}}\sin y,
\end{equation}
where $y=\frac{\sqrt{k}R_{h}}{a}$ and $R_{h}$ is the radius of the event horizon scaled along the $r$ direction
\begin{equation}
R_{h}=a\int_{t}^{\infty}\frac{dt}{a}.\label{Rh}
\end{equation}
We can rewrite Eq.(\ref{ECHDE})
\begin{equation}
\rho_{\Lambda}=3c^{2}M_{p}^{2}L^{-2}\gamma_{\alpha},\label{rrho}
\end{equation}
where 
\begin{equation}
\gamma_{\alpha}=1+\frac{1}{3c^{2}M_{p}^{2}L^{2}}\Big(\alpha \ln(M_{p}^{2}L^{2})+\beta\Big).
\end{equation}
 Eq.(\ref{rrho}) is reduced to (\ref{1}), if $\alpha=\beta=0$. 
 
 In this paper, we will generalize the  LECHDE model
 interacting with Cold Dark Matter (CDM), studied in \cite{setare}. We will determine  the system of first-order differential equations and obtain
the corresponding fixed points, the attractors, repellers and saddle points.
\section{Stability of Interacting Logarithmic Entropy Corrected Holographic Dark Energy Model Solutions}
We suppose that there is an interaction between Logarithmic Entropy Corrected
Holographic Dark Energy (LECHDE) model and dark matter. The continuity equations
yield
\begin{equation}
\dot \rho_{\Lambda}+3H\rho_{\Lambda}(1+\omega_{\Lambda})=-Q,\label{rho}
\end{equation}
\begin{equation}
\dot \rho_{m}+3H\rho_{m}=Q,\label{rho m}
\end{equation}
where $Q=\Gamma \rho_{\Lambda}$ is an interaction term whose form is not
unique. Taking a ratio of two energy densities as $r=\frac{\rho_{m}}{\rho_{\Lambda}}$
and using Eqs.(\ref{rho}) and (\ref{rho m}) we obtain 
\begin{equation}
\dot r=3Hr\Big[\omega_{\Lambda}+\frac{\Gamma}{3H}\frac{(1+r)}{r}\Big].\label{dotr}
\end{equation}
Assuming that there is a transfer from the dark energy component to the matter component, we have $\Gamma>0$ \cite{Bolotin}.  Also, it is clear
from Eq.(\ref{dotr}) that any restriction on the parameter $\omega_{\Lambda}$
will set constraint on the quantity $\Gamma$ \cite{pavon}. Now, we define \cite{kim}
\begin{equation}
\omega_{\Lambda}^{eff}=\omega_{\Lambda}+\frac{\Gamma}{3H},\label{omegaa}
\end{equation}
\begin{equation}
\omega_{m}^{eff}=-\frac{\Gamma}{3Hr}.\label{omegam}
\end{equation}
Using Eqs.(\ref{omegaa}) and (\ref{omegam}), the continuity equations can
be written as
\begin{equation}
\dot \rho_{\Lambda}+3H\rho_{\Lambda}(1+\omega_{\Lambda}^{eff})=0\label{rhodot},
\end{equation}
\begin{equation}
\dot \rho_{m}+3H\rho_{m}(1+\omega_{m}^{eff})=0\label{rhodotm}.
\end{equation}
We assume the non-flat FRW universe 
\begin{equation}
ds^{2}=-dt^{2}+a^{2}(t)\left(\frac{dr^{2}}{1-kr^{2}}+r^{2}d\Omega^{2}\right).
\end{equation}
Here, $k$ denotes the curvature of space with $k=0,1,-1$ for flat, closed
and open universe, respectively. The first Friedmann equation is given by
\begin{equation}
H^{2}+\frac{k}{a^{2}}=\frac{1}{3M_{p}^{2}}\Big[\rho_{\Lambda}+\rho_{m}+\rho_{r}+\rho_{b}\Big]
\label{H},
\end{equation}
where $H$ is the Hubble parameter and $\rho_{\Lambda}, \rho_{m}, \rho_{r}$
and $\rho_{b}$ correspond to the dark energy, CDM, radiation and baryons
densities, respectively. Also, we assume that baryons and radiation have no interaction with dark energy, so they obey the continuity equations
\begin{equation}
\dot \rho_{b}+3H\rho_{b}=0\label{rhodotb},
\end{equation}
\begin{equation}
\dot \rho_{r}+4H\rho_{r}=0\label{rhodotr}.
\end{equation}
We define the dimensionless fractional contributions of baryons, radiation,
CDM, ECHDE and curvature as follows
\begin{equation}
\Omega_{b}=\frac{\rho_{b}}{3M_{p}^2H^{2}},
\end{equation}
\begin{equation}
\Omega_{r}=\frac{\rho_{r}}{3M_{p}^2H^{2}},
\end{equation}
\begin{equation}
\Omega_{m}=\frac{\rho_{m}}{3M_{p}^2H^{2}},
\end{equation}
\begin{equation}
\Omega_{\Lambda}=\frac{\rho_{\Lambda}}{3M_{p}^2H^{2}},\label{oom}
\end{equation}
\begin{equation}
\Omega_{k}=\frac{k}{a^2H^{2}},
\end{equation}
and consider the decay rate to be \cite{pavon}
\begin{equation}
\Gamma=3b^{2}(1+r)H,\label{decay}
\end{equation}
where $b^{2}$ is a coupling constant. Using Eqs. (\ref{omegaa}), (\ref{omegam})
and  (\ref{decay}), we obtain
\begin{equation}
\omega_{\Lambda}^{eff}=\omega_{\Lambda}+b^{2}(\frac{\Omega_{\Lambda}+\Omega_{m}}{\Omega_{\Lambda}}),\label{om}
\end{equation}
\begin{equation}
\omega_{m}^{eff}=-b^{2}(\frac{\Omega_{\Lambda}+\Omega_{m}}{\Omega_{m}}).
\end{equation}
 Taking the time derivative of Eq.(\ref{ECHDE}) and
using Eqs.(\ref{L}), (\ref{Rh}),  (\ref{rho}) and (\ref{decay}), we obtain
\begin{equation}
\omega_{\Lambda}=-1-\frac{b^{2}(\Omega_{\Lambda}+\Omega_{m})}{\Omega_{\Lambda}}+\frac{2}{3}\Big[1-\frac{\sqrt{\Omega_{\Lambda}}
\cos y}{c\sqrt{\gamma_{\alpha}}}\Big]\Big[1+\frac{\gamma_{\alpha}-1}{\gamma_{\alpha}}-\frac{\alpha
H^{2}\Omega_{\Lambda}}{3c^{4}M_{p}^{2}\gamma_{\alpha}^{2}}
\Big],\label{effff}
\end{equation}
which for the non-interacting case  $b^{2}=0$ reduces to
\begin{equation}
\omega_{\Lambda}^{non-int}=-1+\frac{2}{3}\Big[1-\frac{\sqrt{\Omega_{\Lambda}}
\cos y}{c\sqrt{\gamma_{\alpha}}}\Big]\Big[1+\frac{\gamma_{\alpha}-1}{\gamma_{\alpha}}-\frac{\alpha
H^{2}\Omega_{\Lambda}}{3c^{4}M_{p}^{2}\gamma_{\alpha}^{2}}
\Big].
\end{equation}
Now, inserting Eq. (\ref{effff}) in (\ref{om}), we obtain
\begin{equation}
\omega_{\Lambda}^{eff}=-1+\frac{2}{3}\Big[1-\frac{\sqrt{\Omega_{\Lambda}}
\cos y}{c\sqrt{\gamma_{\alpha}}}\Big]\Big[1+\frac{\gamma_{\alpha}-1}{\gamma_{\alpha}}-\frac{\alpha
H^{2}\Omega_{\Lambda}}{3c^{4}M_{p}^{2}\gamma_{\alpha}^{2}}
\Big],\label{eff}
\end{equation}
where the $b^2$ term is canceled out. Taking $\alpha=\beta=0$ in Eq.(\ref{eff})  we recover Eq.(31) in paper
\cite{setare1}.
Also, using Eqs.(\ref{rrho}) and (\ref{oom}) we obtain
\begin{equation}
\Omega_{\Lambda}=\frac{c^{2}\gamma_{\alpha}}{H^{2}L^{2}}.
\end{equation}
Now, we obtain the solutions of the model under investigation and determine their
stability. Following \cite{setare}, we use the quantity 
\begin{equation}
D=\sqrt{H^{2}+\frac{k}{a^{2}}}\label{D},
\end{equation}
and define the dimensionless variables
\begin{equation}
Z=\frac{H}{D}\label{Z},
\end{equation}
\begin{equation}
\widehat{\Omega_{\Lambda}}=\frac{\rho_{\Lambda}}{3M_{p}^2D^{2}}\label{ol},
\end{equation}
\begin{equation}
\widehat{\Omega_{m}}=\frac{\rho_{m}}{3M_{p}^2D^{2}}\label{mm},
\end{equation}
\begin{equation}
\widehat{\Omega_{r}}=\frac{\rho_{r}}{3M_{p}^2D^{2}}\label{oo},
\end{equation}
\begin{equation}
\widehat{\Omega_{b}}=\frac{\rho_{b}}{3M_{p}^2D^{2}}\label{or}.
\end{equation}
 Then, the Friedmann equation takes the following form
\begin{equation}
\widehat{\Omega_{\Lambda}}+\widehat{\Omega_{m}}+\widehat{\Omega_{r}}+\widehat{\Omega_{b}}=1.
\end{equation}
Taking time derivative of Eq.(\ref{H}) and using (\ref{rhodot}),  (\ref{rhodotm}), (\ref{rhodotb}), (\ref{rhodotr}), (\ref{D}) and  (\ref{Z}), we obtain the
deceleration parameter
\begin{equation}
q=-\frac{\dot H}{H^{2}}-1=
\frac{3\Big[\widehat{\Omega_{\Lambda}}\omega_{\Lambda}^{eff}+\widehat{\Omega_{m}}\omega_{m}^{eff}\Big]+\widehat{\Omega_{r}}+1}{2Z^{2}}\label{q}.
\end{equation}
Taking time derivative of Eqs.(\ref{D}), (\ref{Z}), (\ref{ol}), (\ref{mm}),
(\ref{oo}),  (\ref{or}) and using (\ref{rhodot}), (\ref{rhodotm}), (\ref{rhodotb}),
(\ref{rhodotr}), (\ref{Z}), (\ref{ol}), (\ref{oo}), (\ref{or}), 
(\ref{ddt}), (\ref{q}), (\ref{dd}), we obtain
\begin{equation}
\acute D=-Z^{3}D(q+\frac{1}{Z^{2}}),\label{dd}
\end{equation}
\begin{equation}
\acute Z=Z^{2}\Big[-1-q+Z^{2}(q+\frac{1}{Z^{2}})\Big],\label{z}
\end{equation}
\begin{equation}
\acute{\widehat{\Omega_{\Lambda}}}
=\widehat{\Omega_{\Lambda}} Z \Big[-3(1+\omega_{\Lambda}^{eff})+2Z^{2}(q+\frac{1}{Z^{2}})\Big],\label{omega1}
\end{equation}
\begin{equation}
\acute{\widehat{\Omega_{m}}}=
=\widehat{\Omega_{m}}
 Z \Big[-3(1+\omega_{m}^{eff})+2Z^{2}(q+\frac{1}{Z^{2}})\Big],\label{omega2}
\end{equation}
\begin{equation}
\acute{\widehat{\Omega_{r}}}=
=\widehat{\Omega_{r}}
 Z \Big[-4+2Z^{2}(q+\frac{1}{Z^{2}})\Big],\label{omega4}
\end{equation}
\begin{equation}
\acute{\widehat{\Omega_{b}}}=
=\widehat{\Omega_{b}}
 Z \Big[-3+2Z^{2}(q+\frac{1}{Z^{2}})\Big],\label{omega3}
\end{equation}
respectively, where $\acute {}=
\frac{1}{D}\frac{d}{dt}\label{ddt}$. Here $-1\leq Z \leq 1$ and the Hubble parameter can be positive for an expanding
cosmological model or negative for a contracting cosmological model. We study
the dynamical system for the variables $\widehat{\Omega}\equiv(Z,\widehat{\Omega_{\Lambda}},\widehat{\Omega_{m}}
,\widehat{\Omega_{b}},\widehat{\Omega_{r}})$, defined by the Eqs.(\ref{z}),
(\ref{omega1}), (\ref{omega2}), (\ref{omega3}), (\ref{omega4}). The dynamical
character of this system of equations with their fixed points is determined
by the corresponding matrix of  linearization. The real parts
of its eigenvalues will tell us that the cosmological solutions are repeller,
attractor or saddle points \cite{shtanov}.  For $Q\neq0$
and Q=0, the eigenvalues of dynamical system are given in  table 1 and table
4, respectively.

The cosmological models denoted by $DE_{+}$ and $DE_{-}$ are the dark energy
dominated expanding $(H>0)$ and contracting $(H<0)$  models, 
respectively. The cosmological models denoted by $DM_{+}$ and $DM_{-}$ are the matter dominated expanding  and contracting models , respectively.
 The cosmological models denoted by $R_{+}$ and $R_{-}$ are the expanding  and contracting radiation dominated models,  respectively.
 The cosmological models denoted by $B_{+}$ and $B_{-}$ are the baryon dominated
expanding  and contracting models, respectively.
 The cosmological model denoted by $E$ is the Einstein universe $(H=0)$ 
{and
 $M_{+}$ and $M_{-}$ are the expanding  and contracting matter-baryon dominated  models,  respectively.}

The dynamical property of the fixed point is defined by the sign of the
real part of the eigenvalues. If all of the eigenvalues are positive, the point
is said to be a repeller; if all of the eigenvalues are negative, the point
is said to be an attractor; otherwise the fixed point is called a saddle point. For $Q\neq0$,
in table 2 and table 3, we  determine the attractor, repeller and  saddle point characters  for the fixed points given in table 1; and for $Q=0$,
in table 5 and table 6, we  determine the attractor, repeller and  saddle point characters  for the fixed points given in table 4.

\newpage
\hspace{3mm}{\small {\bf Table 1.}} {\small
Fixed points and  eigenvalues for $Q\neq0$.}\\
    \begin{tabular}{l l l l l p{0.15mm} }
  
    \hline\hline
  \vspace{0.50mm}
{\footnotesize  $(Model) $ } & {\footnotesize  $Coordinates$ } & 
{\footnotesize  $~~~~~~~~~~~~~~~~~~~~~~~Eigenvalues$ } \\\hline

\vspace{0.5mm}

{\footnotesize $\big(DE_{+}\big)$}& {\footnotesize $(1,1,0,0,0)$} &
{\footnotesize $1+3\omega_{\Lambda}^{eff},3\omega_{\Lambda}^{eff},3(\omega_{\Lambda}^{eff}-\omega_{m}^{eff})
,-1+3\omega_{\Lambda}^{eff}$}\\

\vspace{0.5mm}

{\footnotesize $\big(DM_{+}\big)$}& {\footnotesize $(1,0,1,0,0)$} &
{\footnotesize $1+3\omega_{m}^{eff},3\omega_{m}^{eff},3(\omega_{m}^{eff}-\omega_{\Lambda}^{eff})
,-1+3\omega_{m}^{eff}$}\\ 

\vspace{0.5mm}

{\footnotesize $\big(R_{+}\big)$}& {\footnotesize $(1,0,0,1,0)$} &
{\footnotesize $1,2,1-3\omega_{\Lambda}^{eff},1-3\omega_{m}^{eff}$}\\

\vspace{0.5mm}

{\footnotesize $\big(B_{+}\big)$}& {\footnotesize $(1,0,0,0,1)$} &
{\footnotesize $-1,0,1,-3\omega_{\Lambda}^{eff},-3\omega_{m}^{eff}$}\\

\vspace{0.5mm}

{\footnotesize $\big(E\big)$}& {\footnotesize $(0,\frac{(-3\widehat{\Omega_{m}}-3\widehat{\Omega_{r}}+3)\omega_{\Lambda}^{eff}+3
\widehat{\Omega_{m}}\omega_{m}^{eff}+\widehat{\Omega_{r
}}+1}{3\omega_{\Lambda}^{eff}},\widehat{\Omega_{m}},\widehat{\Omega_{r}},
-\frac{3\widehat{\Omega_{m}}\omega_{m}^{eff}+\widehat{\Omega_{r}}+1}{3\omega_{\Lambda}^{eff}})$} &
{\footnotesize $~~~~~~~~~~~~~~0$}\\

\vspace{0.5mm}

{\footnotesize $\big(DE_{-}\big)$}& {\footnotesize $(-1,1,0,0,0)$} &
{\footnotesize $-(1+3\omega_{\Lambda}^{eff}),-3\omega_{\Lambda}^{eff},-3(\omega_{\Lambda}^{eff}-\omega_{m}^{eff})
,1-3\omega_{\Lambda}^{eff}$}\\

\vspace{0.5mm}

{\footnotesize $\big(DM_{-}\big)$}& {\footnotesize $(-1,0,1,0,0)$} &
{\footnotesize $-(1+3\omega_{m}^{eff}),-3\omega_{m}^{eff},-3(\omega_{m}^{eff}-\omega_{\Lambda}^{eff})
,1-3\omega_{m}^{eff}$}\\

\vspace{0.5mm}

{\footnotesize $\big(R_{-}\big)$}& {\footnotesize $(-1,0,0,1,0)$} &
{\footnotesize $-1,-2,-1+3\omega_{\Lambda}^{eff},-1+3\omega_{m}^{eff}$}\\

\vspace{0.5mm}

{\footnotesize $\big(B_{-}\big)$}& {\footnotesize $(-1,0,0,0,1)$} &
{\footnotesize $-1,0,1,3\omega_{\Lambda}^{eff},3\omega_{m}^{eff}$} \\  \hline
 \end{tabular}
 \vspace{10mm}


\hspace{3mm}{\small {\bf Table 2.}} {\small
 Attractor, Repeller and  Saddle points for $Q\neq0$.}\\
    \begin{tabular}{l l l l l p{0.15mm} }
    \hline\hline
  \vspace{0.50mm}
{\footnotesize  $(Model) $ } & {\footnotesize ~~~~~~~~~ $Repeller$ } & 
{\footnotesize~~~~~~  $Attractor$ }  & {\footnotesize~~~~~~~~~~~  $Saddle~ point$ } \\\hline

\vspace{0.5mm}

{\footnotesize $\big(DE_{+}\big)$}& {\footnotesize $~~~~~~~~~~~~~\textbf{-}\textbf{-}\textbf{-}\textbf{-}$}&
{\footnotesize $\omega_{\Lambda}^{eff}<\frac{1}{3},\omega_{\Lambda}^{eff}<\omega_{m}^{eff}$}&
{\footnotesize $~~~~~~~~~~~~~~~~~~\textbf{-}\textbf{-}\textbf{-}\textbf{-}$}\\

\vspace{0.5mm}

{\footnotesize $\big(DM_{+}\big)$}& {\footnotesize $~~~~~~~~~~~~~\textbf{-}\textbf{-}\textbf{-}\textbf{-}$} &{\footnotesize $\omega_{m}^{eff}<\frac{1}{3},\omega_{m}^{eff}<\omega_{\Lambda}^{eff}$}&
{\footnotesize $~~~~~~~~~~~~~~~~~~\textbf{-}\textbf{-}\textbf{-}\textbf{-}$}\\ 

\vspace{0.5mm}

{\footnotesize $\big(R_{+}\big)$}& {\footnotesize $\omega_{\Lambda}^{eff}<\frac{1}{3},\omega_{m}^{eff}<\frac{1}{3}$} &{\footnotesize $~~~~~~~~~~~~~\textbf{-}\textbf{-}\textbf{-}\textbf{-}$}
&{\footnotesize $~~~~~~~~~~~~~~~~~~\textbf{-}\textbf{-}\textbf{-}\textbf{-}$}\\

\vspace{0.5mm}

{\footnotesize $\big(B_{+}\big)$}&  {\footnotesize $~~~~~~~~~~~~~\textbf{-}\textbf{-}\textbf{-}\textbf{-}$} & {\footnotesize $~~~~~~~~~~~~~\textbf{-}\textbf{-}\textbf{-}\textbf{-}$}&
{\footnotesize ~~~~~~~~~~~$\textbf{Saddle point}$}\\

\vspace{0.5mm}

{\footnotesize $\big(E\big)$}& {\footnotesize $~~~~~~~~~~~~~\textbf{-}\textbf{-}\textbf{-}\textbf{-}$} & {\footnotesize $~~~~~~~~~~~~~\textbf{-}\textbf{-}\textbf{-}\textbf{-}$}&
{\footnotesize~~~~~~~~~~ $\textbf{Saddle point}$}\\

\vspace{0.5mm}

{\footnotesize $\big(DE_{-}\big)$} &{\footnotesize $\omega_{\Lambda}^{eff}<\frac{1}{3},\omega_{\Lambda}^{eff}<\omega_{m}^{eff}$}&
{\footnotesize $~~~~~~~~~~~~~\textbf{-}\textbf{-}\textbf{-}\textbf{-}$}& {\footnotesize $~~~~~~~~~~~~~~~~~~\textbf{-}\textbf{-}\textbf{-}\textbf{-}$}\\

\vspace{0.5mm}

{\footnotesize $\big(DM_{-}\big)$}& {\footnotesize $\omega_{m}^{eff}<\frac{1}{3},\omega_{m}^{eff}<\omega_{\Lambda}^{eff}$} &{\footnotesize $~~~~~~~~~~~~~\textbf{-}\textbf{-}\textbf{-}\textbf{-}$}
&{\footnotesize $~~~~~~~~~~~~~~~~~~\textbf{-}\textbf{-}\textbf{-}\textbf{-}$}\\

\vspace{0.5mm}

{\footnotesize $\big(R_{-}\big)$}& {\footnotesize $~~~~~~~~~~~~~\textbf{-}\textbf{-}\textbf{-}\textbf{-}$} &{\footnotesize $\omega_{\Lambda}^{eff}<\frac{1}{3},\omega_{m}^{eff}<\frac{1}{3}$}
& {\footnotesize $~~~~~~~~~~~~~~~~~~\textbf{-}\textbf{-}\textbf{-}\textbf{-}$}\\

\vspace{0.5mm}

{\footnotesize $\big(B_{-}\big)$}& {\footnotesize $~~~~~~~~~~~~~\textbf{-}\textbf{-}\textbf{-}\textbf{-}$} & {\footnotesize $~~~~~~~~~~~~~\textbf{-}\textbf{-}\textbf{-}\textbf{-}$}&
{\footnotesize~~~~~~~~~~ $\textbf{Saddle point}$}\\ \hline
 \end{tabular}
 \vspace{10mm}


\hspace{3mm}{\small {\bf Table 3.}} {\small
 Attractor, Repeller and  Saddle points for $Q\neq0$.}\\
    \begin{tabular}{l l l l l p{0.15mm} }
    \hline\hline
  \vspace{0.50mm}
{\footnotesize  $(Model) $ } & {\footnotesize~~~~~~~~~  $Repeller$ } & 
{\footnotesize~~~~~~~~  $Attractor$ }  & {\footnotesize  $Saddle~ point$ } \\\hline

\vspace{0.5mm}

{\footnotesize $\big(DE_{+}\big)$}& {\footnotesize $\omega_{\Lambda}^{eff}>\frac{-1}{3},\omega_{\Lambda}^{eff}>\omega_{m}^{eff}$}&
{\footnotesize $~~~~~~~~~~~~~\textbf{-}\textbf{-}\textbf{-}\textbf{-}$}&
{\footnotesize $~~~~~\textbf{-}\textbf{-}\textbf{-}\textbf{-}$}\\

\vspace{0.5mm}

{\footnotesize $\big(DM_{+}\big)$}& {\footnotesize $\omega_{m}^{eff}>\frac{-1}{3},\omega_{m}^{eff}>\omega_{\Lambda}^{eff}$} &{\footnotesize $~~~~~~~~~~~~~\textbf{-}\textbf{-}\textbf{-}\textbf{-}$}&
{\footnotesize $~~~~~\textbf{-}\textbf{-}\textbf{-}\textbf{-}$}\\

\vspace{0.5mm}

{\footnotesize $\big(DE_{-}\big)$} &{\footnotesize $~~~~~~~~~~~~~\textbf{-}\textbf{-}\textbf{-}\textbf{-}$}&
{\footnotesize $\omega_{\Lambda}^{eff}>\frac{-1}{3},\omega_{\Lambda}^{eff}>\omega_{m}^{eff}$}& {\footnotesize $~~~~~\textbf{-}\textbf{-}\textbf{-}\textbf{-}$}\\

\vspace{0.5mm}

{\footnotesize $\big(DM_{-}\big)$}& {\footnotesize $~~~~~~~~~~~~~\textbf{-}\textbf{-}\textbf{-}\textbf{-}$} &{\footnotesize $\omega_{m}^{eff}>\frac{-1}{3},\omega_{m}^{eff}>\omega_{\Lambda}^{eff}$}
&{\footnotesize $~~~~~\textbf{-}\textbf{-}\textbf{-}\textbf{-}$}\\  \hline

\vspace{0.5mm}
 \end{tabular}
 \vspace{5mm}


\hspace{3mm}{\small {\bf Table 4.}} {\small
Fixed points and  eigenvalues for Q=0.}\\
    \begin{tabular}{l l l l l p{0.15mm} }
  
    \hline\hline
  \vspace{0.50mm}
{\footnotesize  $(Model) $ } & {\footnotesize  $Coordinates$ } & 
{\footnotesize  $~~~~~~~~~~~~~~~~~~~~~~~Eigenvalues$ } \\\hline

\vspace{0.5mm}

{\footnotesize $\big(DE_{+}\big)$}& {\footnotesize $(1,1,0,0,0)$} &
{\footnotesize $1+3\omega_{\Lambda}^{non-int},3\omega_{\Lambda}^{non-int}
,-1+3\omega_{\Lambda}^{non-int}$}\\

\vspace{0.5mm}

{\footnotesize $\big(M_{+}\big)$}& {\footnotesize $(1,0,1-\widehat{\Omega_{b}},\widehat{\Omega_{b}},0)$} &
{\footnotesize $1,0,-1
,-3\omega_{\Lambda}^{non-int}$}\\ 

\vspace{0.5mm}

{\footnotesize $\big(R_{+}\big)$}& {\footnotesize $(1,0,0,1,0)$} &
{\footnotesize $1,2,1-3\omega_{\Lambda}^{non-int}$}\\

\vspace{0.5mm}

{\footnotesize $\big(E\big)$}& {\footnotesize $(0,-\frac{\widehat{\Omega_{r}}+1}{3\omega_{\Lambda}^{non-int}},\frac{-3(\widehat{\Omega_{b}}
+\widehat{\Omega_{r}}-1)\omega_{\Lambda}^{non-int}+\widehat{\Omega_{r}}+1}{3\omega_{\Lambda}^{non-int}}
,\widehat{\Omega_{b}},\widehat{\Omega_{r}}$)} &
{\footnotesize $~~~~~~~~~~~~~~0$}\\

\vspace{0.5mm}

{\footnotesize $\big(DE_{-}\big)$}& {\footnotesize $(-1,1,0,0,0)$} &
{\footnotesize $-(1+3\omega_{\Lambda}^{non-int}),-3\omega_{\Lambda}^{non-int}
,1-3\omega_{\Lambda}^{non-int}$}\\

\vspace{0.5mm}

{\footnotesize $\big(M_{-}\big)$}& {\footnotesize $(-1,0,1-\widehat{\Omega_{b}},\widehat{\Omega_{b}},0)$} &
{\footnotesize $-1,0,1
,3\omega_{\Lambda}^{non-int}$}\\

\vspace{0.5mm}

{\footnotesize $\big(R_{-}\big)$}& {\footnotesize $(-1,0,0,1,0)$} &
{\footnotesize $-1,-2,-1+3\omega_{\Lambda}^{non-int}$}\\

\vspace{0.5mm}
 \\  \hline
 \end{tabular}
 \vspace{10mm}

\hspace{3mm}{\small {\bf Table 5.}} {\small
 Attractor, Repeller and  Saddle points for Q=0.}\\
    \begin{tabular}{l l l l l p{0.15mm} }
    \hline\hline
  \vspace{0.50mm}
{\footnotesize  $(Model) $ } & {\footnotesize ~~~~~~~~~ $Repeller$ } & 
{\footnotesize~~~~~~  $Attractor$ }  & {\footnotesize~~~~~~~~~~~  $Saddle~ point$ } \\\hline

\vspace{0.5mm}

{\footnotesize $\big(DE_{+}\big)$}& {\footnotesize $\omega_{\Lambda}^{non-int}>-\frac{1}{3}$}&
{\footnotesize $~~~~~~~~~\textbf{-}\textbf{-}\textbf{-}\textbf{-}$}&
{\footnotesize $~~~~~~~~~~~~~~~~~~\textbf{-}\textbf{-}\textbf{-}\textbf{-}$}\\

\vspace{0.5mm}

{\footnotesize $\big(M_{+}\big)$}& {\footnotesize $~~~~~~~~~~~~~\textbf{-}\textbf{-}\textbf{-}\textbf{-}$} &{\footnotesize $~~~~~~~~~~\textbf{-}\textbf{-}\textbf{-}\textbf{-}$}&
{\footnotesize $~~~~~~~~~~\textbf{Saddle point}$}\\ 

\vspace{0.5mm}

{\footnotesize $\big(R_{+}\big)$}& {\footnotesize $\omega_{\Lambda}^{non-int}<\frac{1}{3}$} &{\footnotesize $~~~~~~~~~~\textbf{-}\textbf{-}\textbf{-}\textbf{-}$}
&{\footnotesize $~~~~~~~~~~~~~~~~~~\textbf{-}\textbf{-}\textbf{-}\textbf{-}$}\\

\vspace{0.5mm}

{\footnotesize $\big(E\big)$}& {\footnotesize $~~~~~~~~~~~~~\textbf{-}\textbf{-}\textbf{-}\textbf{-}$} & {\footnotesize $~~~~~~~~~~\textbf{-}\textbf{-}\textbf{-}\textbf{-}$}&
{\footnotesize~~~~~~~~~~ $\textbf{Saddle point}$}\\

\vspace{0.5mm}

{\footnotesize $\big(DE_{-}\big)$} &{\footnotesize $~~~~~~~~~~~~~\textbf{-}\textbf{-}\textbf{-}\textbf{-}$}&
{\footnotesize $\omega_{\Lambda}^{non-int}>-\frac{1}{3}$}& {\footnotesize $~~~~~~~~~~~~~~~~~~\textbf{-}\textbf{-}\textbf{-}\textbf{-}$}\\

\vspace{0.5mm}

{\footnotesize $\big(M_{-}\big)$}& {\footnotesize $~~~~~~~~~~~~~\textbf{-}\textbf{-}\textbf{-}\textbf{-}$} &{\footnotesize $~~~~~~~~~~~~~\textbf{-}\textbf{-}\textbf{-}\textbf{-}$}
&{\footnotesize $~~~~~~~~~~~\textbf{Saddle point}$}\\

\vspace{0.5mm}

{\footnotesize $\big(R_{-}\big)$}& {\footnotesize $~~~~~~~~~~~~~\textbf{-}\textbf{-}\textbf{-}\textbf{-}$} &{\footnotesize $\omega_{\Lambda}^{non-int}<\frac{1}{3}$}
& {\footnotesize $~~~~~~~~~~~~~~~~~~\textbf{-}\textbf{-}\textbf{-}\textbf{-}$}\\

\vspace{0.5mm}

\\ \hline
 \end{tabular}
 \vspace{10mm}


\hspace{3mm}{\small {\bf Table 6.}} {\small
 Attractor, Repeller and  Saddle points for Q=0.}\\
    \begin{tabular}{l l l l l p{0.15mm} }
    \hline\hline
  \vspace{0.50mm}
{\footnotesize  $(Model) $ } & {\footnotesize~~~~~~~~~  $Repeller$ } & 
{\footnotesize~~~~~~~~  $Attractor$ }  & {\footnotesize  $Saddle~ point$ } \\\hline

\vspace{0.5mm}

{\footnotesize $\big(DE_{+}\big)$}& {\footnotesize $~~~~~~~~~~~~~\textbf{-}\textbf{-}\textbf{-}\textbf{-}$}&
{\footnotesize $\omega_{\Lambda}^{non-int}<\frac{1}{3}$}&
{\footnotesize $~~~~~\textbf{-}\textbf{-}\textbf{-}\textbf{-}$}\\

\vspace{0.5mm}

{\footnotesize $\big(DE_{-}\big)$} &{\footnotesize $\omega_{\Lambda}^{non-int}<\frac{1}{3}$}&
{\footnotesize $~~~~~~~~~~~~~\textbf{-}\textbf{-}\textbf{-}\textbf{-}$}& {\footnotesize $~~~~~\textbf{-}\textbf{-}\textbf{-}\textbf{-}$}\\
\\  \hline
 \end{tabular}

\section{Concluding remarks}\label{Con}
In this work, we have discussed the cosmological dynamics of interacting Logarithmic Entropy Corrected Holographic Dark Energy model. We have determined
the system of first-order differential equations that explains the evolution
of the five dimensionless quantities. In addition, for $Q\neq0$, the nine fixed points
of the mentioned cosmological model are obtained and the dynamical properties
of these fixed points are presented. Also, for $Q=0$, the seven fixed points
of the mentioned cosmological model are obtained and the dynamical properties
of these fixed points are presented.

In particular, for $Q\neq0$, it is shown that  the Dark Energy dominated models
($DE_{+}$ and $DE_{-}$) have a set of attractor and repeller points. Considering the conditions $\omega_{\Lambda}^{eff}<\frac{1}{3}$
and $\omega_{\Lambda}^{eff}<\omega_{m}^{eff}$, we have shown that  the expanding
 Dark Energy  dominated model ($DE_{+}$) and the contracting Dark Energy  dominated model ($DE_{-}$) are  attractor and repeller, respectively. Similarly,  considering the conditions $\omega_{m}^{eff}<\frac{1}{3}$
and $\omega_{m}^{eff}<\omega_{\Lambda}^{eff}$, we have shown that  the expanding
Dark Matter dominated model ($DM_{+}$) and the contracting Dark Matter dominated model ($DM_{-}$) are  attractor and repeller, respectively. Finally,
considering the conditions $\omega_{\Lambda}^{eff}<\frac{1}{3}$
and $\omega_{m}^{eff}<\frac{1}{3}$, we have shown that  the expanding
Radiation dominated model ($R_{+}$) and the contracting Radiation dominated
model ($R_{-}$) are repeller  and attractor, respectively. We have shown that the expanding early universe model ($B_{+}$), the contracting early universe  model ($B_{-}$)  and  the Einstein universe model (E) are saddle points.

Also, for $Q=0$, it is shown that  the Dark Energy dominated models
($DE_{+}$ and $DE_{-}$) have a set of attractor and repeller points. Considering the conditions $\omega_{\Lambda}^{non-int}<\frac{1}{3}$, we have shown that  the expanding
 Dark Energy  dominated model ($DE_{+}$) and the contracting Dark Energy  dominated model ($DE_{-}$) are  attractor and repeller, respectively. Similarly,  considering the conditions $\omega_{\Lambda}^{non-int}<\frac{1}{3}$, we have shown that  the expanding
radiation dominated model ($R_{+}$) and the contracting radiation dominated model ($R_{-}$) are  repeller and attractor, respectively. Finally,
we have shown that the expanding matter-baryon universe model ($M_{+}$), the contracting matter-baryon universe  model ($M_{-}$)  and  the Einstein universe model (E) are saddle points.


\end{document}